\documentclass[twocolumn,showpacs,preprintnumbers,amsmath,amssymb,superscriptaddress, prl,floatfix]{revtex4-1}
\usepackage{bm}

\usepackage{graphicx}
\usepackage{dcolumn}
\usepackage{tabularx}
\usepackage{bm}
\usepackage{epstopdf}
\usepackage{amsmath}
\usepackage{color}
\usepackage{here}
\begin{document}

\preprint{Preprint}

\title{Dephasing effects on coherent exciton-polaritons and the breaking of the strong coupling regime}

\author{N. Takemura}
\email[E-mail: ]{naotomo.takemura@epfl.ch}
\affiliation{Laboratory of Quantum Optoelectronics, \'Ecole Polytechnique F\'ed\'erale de Lausanne, CH-1015, Lausanne, Switzerland}
\author{S. Trebaol}
\affiliation{UMR FOTON, CNRS, Universit\'e de Rennes 1, Enssat, F22305 Lannion, France.}
\author{M. D. Anderson}
\affiliation{Laboratory of Quantum Optoelectronics, \'Ecole Polytechnique F\'ed\'erale de Lausanne, CH-1015, Lausanne, Switzerland}
\author{S. Biswas}
\affiliation{Laboratory of Quantum Optoelectronics, \'Ecole Polytechnique F\'ed\'erale de Lausanne, CH-1015, Lausanne, Switzerland}
\author{ D. Y. Oberli}
\affiliation{Laboratory of Quantum Optoelectronics, \'Ecole Polytechnique F\'ed\'erale de Lausanne, CH-1015, Lausanne, Switzerland}
\author{M. T. Portella-Oberli}
\affiliation{Laboratory of Quantum Optoelectronics, \'Ecole Polytechnique F\'ed\'erale de Lausanne, CH-1015, Lausanne, Switzerland}
\author{B. Deveaud}
\affiliation{Laboratory of Quantum Optoelectronics, \'Ecole Polytechnique F\'ed\'erale de Lausanne, CH-1015, Lausanne, Switzerland}

\date{\today}

\begin{abstract}
Using femtosecond pump-probe spectroscopy, we identify excitation induced dephasing as a major mechanism responsible for the breaking of the strong-coupling between excitons and photons in a semiconductor microcavity. The effects of dephasing are observed on the transmitted probe pulse spectrum as a density dependent broadening of the exciton-polariton resonances and the emergence of a third resonance at high excitation density. A striking asymmetry in the energy shift between the upper and the lower polaritons is also evidenced. Using the excitonic Bloch equations, we quantify the respective contributions to the energy shift of many-body effects associated with Fermion exchange and photon assisted exchange processes and the contribution to collisional broadening.
\end{abstract}


\maketitle
A semiconductor microcavity is a system that confines photons and allows them to strongly interact with quantum well excitons \cite{Weisbuch1992}. Polaritons are composite particles arising from the coherent superposition of a photon and exciton. The interactions mediated by their excitonic part make semiconductor microcavities a suitable platform for realizing nonlinear optical devices such as: bistablity memory \cite{Cerna2013} and polariton switching \cite{Amo2010}. Moreover, applications of semiconductor microcavities to quantum information have been also proposed on the basis of their coherent behaviour \cite{Demirchyan2014,Dominici2014}. When modeling these devices, they are usually described with a nonlinear Schr\"{o}dinger equation, which is formally the same as the Gross-Pitaevskii equations (GPE) used for coherent ground-state of Bose condensed dilute atoms. This directly leads to a wide range of analogies between exciton-polaritons and cold atoms. Actually, in semiconductor microcavity systems, we can investigate a wide range of physics including Bose-Einstein condensation \cite{Kasprzak2006,Deng2002} and the collective quantum fluid nature \cite{Amo2009,Kohnle2011, Kohnle2012}. On the other hand, one of the important properties of a semiconductor system is the existence of dephasing \cite{Shah1999,Kira2006,Wang1998}, which induces decoherence. The investigation of the effect of decoherence, induced by excitonic dephasing, on the polariton dynamics is important both for understanding the physics of exciton-photon strongly coupled systems and for designing semiconductor microcavity devices such as nonlinear optical devices and polariton-based qubits.
  
In this letter, we show that excitation induced dephasing (EID) plays an important role in the dynamics of polaritons in a semiconductor microcavity. The investigation is experimentally carried out by time-resolved femtosecond pump-probe optical spectroscopy. For the theoretical description of our results, we utilize the excitonic Bloch equations (EBE) approach, taking into account separately the coherent part of the polariton population and an incoherent population of excitons \cite{Rochat2000,Wang2009}. We study the role of exciton-exciton interactions, photon-assisted exchange scattering, and EID effects on the polariton dynamics. The experimental results are very well reproduced by EBE and not by the exciton-photon GPE, which assumes that excitons are in a coherent limit.

The experiment is performed in a high quality GaAs-based microcavity \cite{Stanley1994} at the cryogenic temperature of 4K. A single 8 nm In$_{0.04}$Ga$_{0.96}$As quantum well is embedded in between two GaAs/AlAs distributed Bragg-reflectors. The Rabi splitting energy is $2\Omega$=3.45 meV at zero cavity detuning \cite{Kohnle2011}. For the accurate measurement of the transmitted probe beam, we employ a heterodyne pump-probe setup with a degenerated beam configuration at $\bm k=0$ $\mu$m$^{-1}$ \cite{Takemura2014a}, which dramatically increases the signal-to-noise ratio. The pump and probe pulses originate both from a broadband few hundreds femtosecond Ti:Sapphire laser. The center of the laser spectrum is set between the lower and upper-polariton peaks. Additionally, noise coming from laser spectrum envelope is removed with the aid of a numerical low-pass filter. The experimental setup is described in detail in our previous papers \cite{Kohnle2012,Takemura2014a}. In order to avoid the complex effects of biexcitons \cite{Takemura2014,Takemura2014a}, the pump and probe beams are co-circularly polarized. We obtain a time delay map through successive measurements of the pump-probe spectrum. 

\begin{figure}
\includegraphics[width=0.441\textwidth]{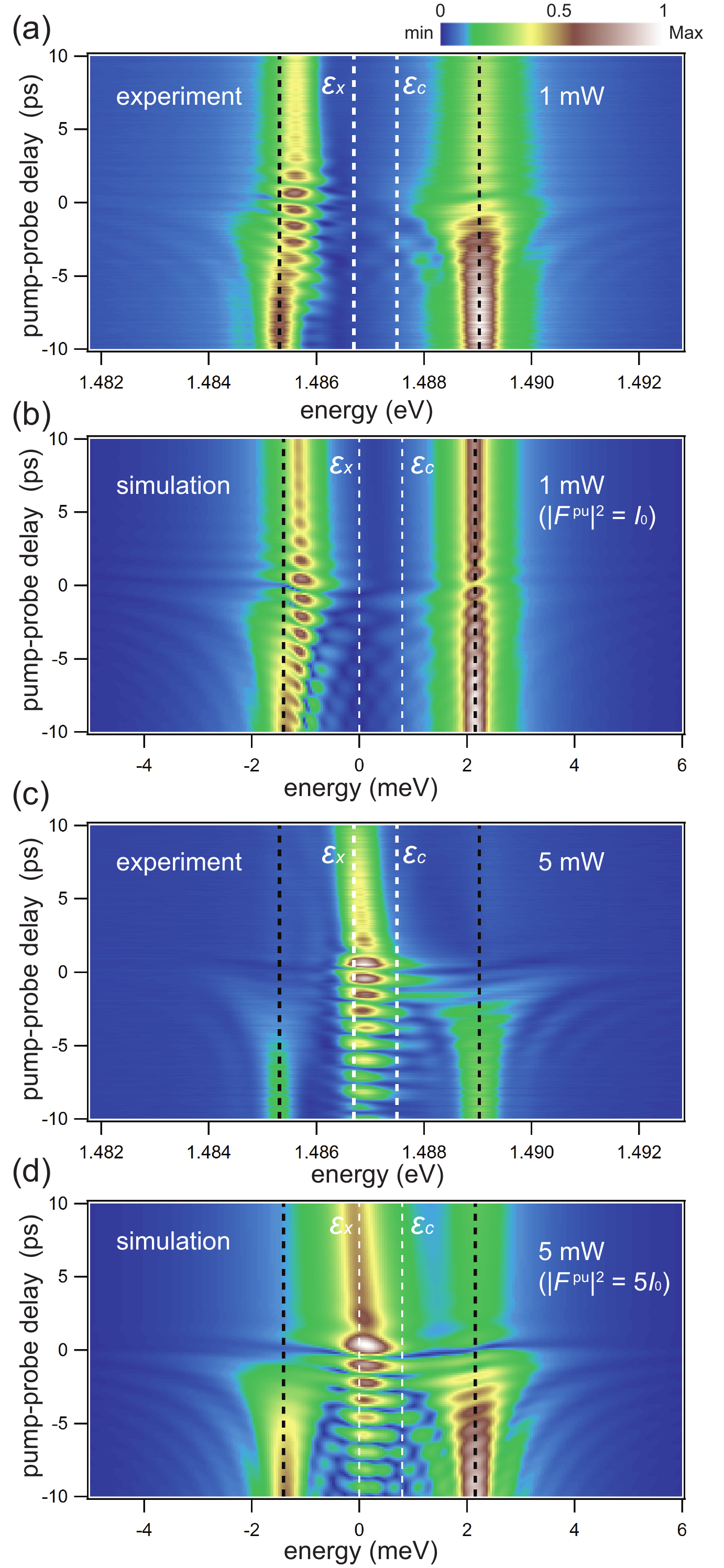}
\caption{(color online) Measured and simulated probe transmission are shown as a function of energy and time delay between pump and probe pulse. The spectra are measured for two different pump intensities: 1.48$\times$10$^{13}$ (1 mW) (a) and 7.4$\times$10$^{13}$ (5 mW) (c) photons pulse$^{-1}$ cm$^{-2}$. Corresponding simulated spectra are attached below the measured spectra ((b) and (c)). The black dashed lines represent the lower and upper-polariton peak energies without pump pulse. The white dashed lies are the cavity photon and exciton energies. In the simulation, the intensity $I_0$ is defined as $I_0=0.8/g_0$.}
\label{fig:fig1}
\end{figure}

The experimental results are presented in Fig. \ref{fig:fig1} (a) and (c), showing the probe spectra as a function of pump-probe time delay. The cavity detuning is set at $\epsilon_{c}-\epsilon_{x}=0.8$ meV, where $\epsilon_{c(x)}$ is cavity mode (exciton) energy. In this figure, the pump pulse arrives before (after) the probe pulse at positive (negative) pump-probe delays. For low pump intensity (Fig. \ref{fig:fig1} (a)), we observe two polariton branches (lower and upper) at both positive and negative delays and the lower polariton shows a maximum blue-shift at zero delay. The delay dependence of the lower-polariton blue shift is asymmetric with respect to zero delay. While the blue shift gradually decreases at negative delays, it stays at positive delays for long time. No clear energy shift of the upper polariton resonance is seen for all delays. At a high pump intensity (Fig. \ref{fig:fig1} (c)), a triple peak structure appears at negative delays, while a single peak exists at positive delays. With the aid of numerical simulations based on EBE, we show that such behaviours originate from a long-living incoherent population and a short-living coherent polarization of excitons.   

For the analysis of the experiment, we use EBE \cite{Rochat2000,Wang2009}. The starting point of EBE is a bosonic exciton-photon Hamiltonian:
\begin{eqnarray}
\hat{H}&=&\int d{\bf x} \left[\epsilon_{x}\hat{\bm \psi}_{x}^{\dagger}\hat{\bm \psi}_{x}+\hat{\bm \psi}_{c}^{\dagger}(\epsilon_{c}-\frac{\hbar^2\nabla^2}{2m_{c}})\hat{\bm \psi}_{c}\right.\nonumber\\
& &\left.+\Omega(\hat{\bm \psi}_{c}^{\dagger}\hat{\bm \psi}_{x}+\hat{\bm \psi}_{x}^{\dagger}\hat{\bm \psi}_{c})\right]\nonumber\\
& &+\int d{\bf x}\int d{\bf x'}\left[\frac{1}{2}\hat{\bm \psi}_{x}^{\dagger}\hat{\bm \psi}_{x}^{\dagger\prime}V_{\rm ex}({\bf x}-{\bf x'})\hat{\bm \psi}_{x}\hat{\bm \psi}_{x}'\right.\\
& &\left.-V_{\rm pae}({\bf x}-{\bf x'})(\hat{\bm \psi}_{c}^{\dagger}\hat{\bm \psi}_{x}^{\dagger\prime}\hat{\bm \psi}_{x}\hat{\bm \psi}_{x}'+\hat{\bm \psi}_{x}^{\dagger}\hat{\bm \psi}_{x}^{\dagger\prime}\hat{\bm \psi}_{x}\hat{\bm \psi}_{c}')\right].\nonumber
\end{eqnarray}
$\hat{\bm \psi}_{x(c)}$ and $\hat{\bm \psi}_{x(c)}^{\dagger}$ are exciton (photon) field creation and annihilation operators. They obey the Bose commutation relation, $[\hat{\bm \psi}_{x(c)},\hat{\bm \psi}_{x(c)}^{\dagger\prime}]=\delta({\bf x}-{\bf x'})$. This Hamiltonian can be obtained from the electron-hole Hamiltonian via a boson mapping method called the Usui transformation \cite{Rochat2000}. Since the exciton mass is large, the kinetic term of the exciton is neglected. The interactions are assumed to be contact interactions: $V_{\rm ex}({\bf x}-{\bf x'})=g\delta({\bf x}-{\bf x'})$ and $V_{\rm pae}({\bf x}-{\bf x'})=g_{\rm pae}\delta({\bf x}-{\bf x'})$ \cite{pitaevskii2003bose,Carusotto2013}. The exciton-exciton interaction potential $V_{\rm ex}$ is associated with the Coulomb exchange scattering. The term $V_{\rm pae}$ is a photon assisted exchange scattering \cite{Combescot2007} and contributes to the reduction of the Rabi coupling, which is the reminiscence of the fermionic nature of the exciton \cite{Rochat2000}. In order to obtain a closed set of equations, we truncate the hierarchy by applying the following assumptions such as $\langle\hat{\bm \psi}_{x}^{\dagger}\hat{\bm \psi}_{x}\hat{\bm \psi}_{x}\rangle\simeq\langle\hat{\bm \psi}_{x}^{\dagger}\hat{\bm \psi}_{x}\rangle\langle\hat{\bm \psi}_{x}\rangle$, $\langle\hat{\bm \psi}_{x}^{\dagger\prime}\hat{\bm \psi}_{x}'\hat{\bm \psi}_{x}\rangle\simeq\langle\hat{\bm \psi}_{x}^{\dagger\prime}\hat{\bm \psi}_{x}'\rangle\langle\hat{\bm \psi}_{x}\rangle$, $\langle\hat{\bm \psi}_{x}\hat{\bm \psi}_{x}\rangle=0$, and $\langle\hat{\bm \psi}_{x}^{\dagger}\hat{\bm \psi}_{x}'\hat{\bm \psi}_{x}'\rangle\simeq\langle\hat{\bm \psi}_{x}^{\dagger}\rangle\langle\hat{\bm \psi}_{x}'\hat{\bm \psi}_{x}'\rangle=0$. We define the exciton population as $N({\bf x},t)=\langle\hat{\bm \psi}_{x}^{\dagger}\hat{\bm \psi}_{x}\rangle$ and the exciton polarization as $P({\bf x},t)=\langle\hat{\bm \psi}_{x}\rangle$. Assuming factorization between the photon and exciton, we define $E({\bf x},t)=\langle\hat{\bm \psi}_{c}\rangle$. With the aid of the Heisenberg equation of motion, the EBE then reads \cite{Rochat2000}
\begin{eqnarray}
i\hbar \dot{N}&=&-i\Gamma_x N-2i(\Omega-2g_{\rm pae}N){\rm Im}[PE^{*}]\nonumber\\
i\hbar \dot{P}&=&\left(\epsilon_x+g_0N-i\gamma_x(N)\right)P+(\Omega-2g_{\rm pae}N)E\\
i\hbar \dot{E}&=&(\epsilon_c-\frac{\hbar^2}{2m_{c}}\nabla^2-i\gamma_{c})E+(\Omega-g_{\rm pae}N)P - f_{\rm ext}.\nonumber
\end{eqnarray}
To obtain the above equations, the interaction constant $g$ is phenomenologically divided into a real and imaginary part: $g=g_0-ig'$. The real part $g_0$ is associated with an energy renormalization, while  the imaginary part $g'$ represents the strength of EID, which is also referred to as collisional broadening. The ratio of the constants is estimated as $g_{\rm pae}/g_0\simeq\hbar\Omega/6n_sE_ba_0^2$ \cite{Sarchi2007}. $n_s$ is the saturation density of excitons. $E_b$ and $a_0$ are respectively the exciton binding energy and Bohr radius. The constants, $\gamma_x(N)$ and $\Gamma_x$ are respectively polarization dephasing and population decay rate of excitons. In general, $\gamma_x(N)$ is written as \cite{Shah1999,Baars2000,Ciuti1998},
\begin{equation}
\gamma_x(N)=\Gamma_x/2+\gamma_x^{*}+g'N,
\end{equation} 
where $\gamma^{*}_x$ is the pure dephasing term. In the terminology of two level systems, $\Gamma_x$ and $\gamma_x$ correspond to the inverse of $T_1$ and $T_2$ times respectively . The EID constant $g'$ introduces a phenomenological linear increase of the dephasing that depends on the exciton population $N$, which plays an important role in our experiment. 
 
The advantage of EBE, compared to GPE, is that we can apply independent decay rates for the coherent polarization and incoherent population and calculate the time evolution of each. Indeed, in the commonly used GPE, a factorization,    
$\langle\hat{\bm \psi}_{x}^{\dagger}\hat{\bm \psi}_{x}\rangle=\langle\hat{\bm \psi}_{x}^{\dagger}\rangle\langle\hat{\bm \psi}_{x}\rangle$ is implicitly assumed \cite{Carusotto2013}. Therefore, the dynamics of the population $N({\bf x},t)$ is uniquely determined by the polarization $P({\bf x},t)$ through $N=|P|^2$. In the EBE, this condition is satisfied when neither pure dephasing nor EID exist: $\gamma_x^*=0$ meV and $g'=0$ meV (coherent limit).
The EBE are close analogue of the optical Bloch equations (OBE) \cite{Li2006,Wang1998}, however differing since OBE are based on a two-level electron-hole system, while EBE are based on a bosonic exciton basis \cite{Rochat2000}. 

To reproduce the experiments, $\Gamma_x$ and $\gamma_c$ are chosen to be 0.01 meV and 0.1 meV  respectively. The pure dephasing is set to $\gamma_x^*=0.1$ meV \cite{Honold1989}, additionally, we include EID as the primary decoherence mechanism in our simulations. We set the interaction constants as $g'=0.4g_0$ and $g_{\rm pae}=0.3g_0$. $f_{\rm ext}$ is the excitation photon field and is assumed to be a Gaussian pulse: $f_{\rm ext}=F^{pu(pr)}\exp(-(t-t_0)^2/(\tau^2))\exp(-i\omega_0^{\rm pulse}t)$. We set $\omega_0^{\rm pulse}$ at the center of both polariton branches and set $\tau$=0.5 ps.

For the calculation of the pump-probe dynamics, we use a coupled-mode approximation: $N({\bf x},t)=N^{pu}+N^{pr}e^{i{\bm k}\cdot{\bf x}}+N^{pr*}e^{-i{\bm k}\cdot{\bf x}}$ (the population is a real value), $P({\bf x},t)=P^{pu}+P^{pr}e^{i{\bm k}\cdot{\bf x}}+P^{id}e^{-i{\bm k}\cdot{\bf x}}$, and $E({\bf x},t)=E^{pu}+E^{pr}e^{i{\bm k}\cdot{\bf x}}+E^{id}e^{-i{\bm k}\cdot{\bf x}}$. For example, $P^{pu}$, $P^{pr}$, and $P^{id}$ represent pump, probe, and idler component of the polarization respectively. Considering the conservation of momentum, we obtain 8 coupled equations (See Supplementary material). The pump and probe pulses are introduced as $E^{\rm pu}$ and $E^{\rm pr}$ respectively and the transmitted probe signal is obtained through $E^{\rm pr}$.  This is the standard method of calculating a transient four-wave mixing signal in optical Bloch equations \cite{Yajima1979,Turner2011}. Since the wave number of the probe is sufficiently small, we neglect the momentum dispersion of the photon mode.         
\begin{figure}
\includegraphics[width=0.441\textwidth]{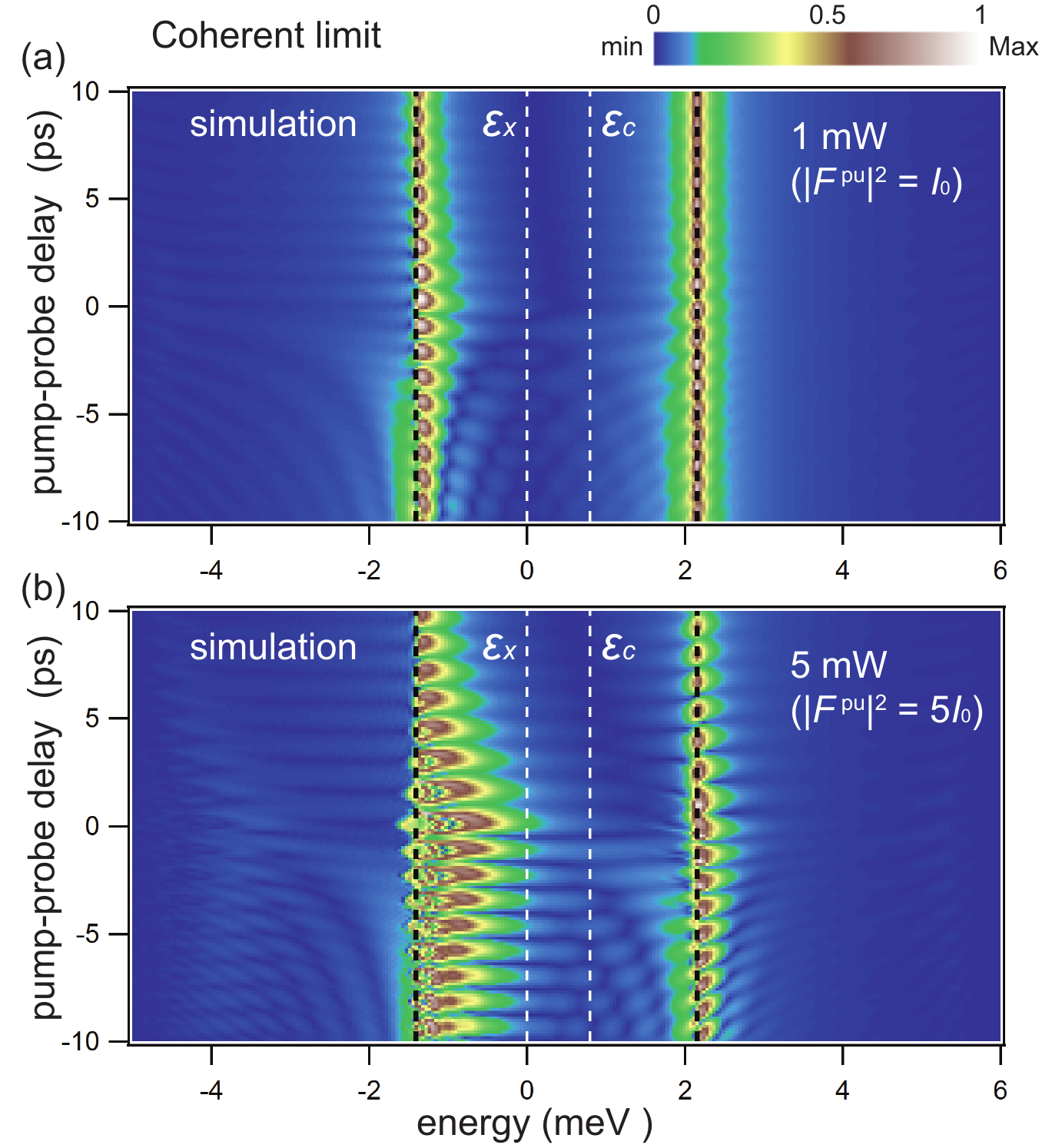}
\caption{(color online) Simulated probe transmission as a function of energy and pump-probe time delay without EID or pure dephasing ($g'=\gamma_x^*=0$ meV). The other parameters are the same as those used in the simulation of Fig. 1. Fig. 2 (a) and (b) respectively correspond to 1 mW and 5 mW pump intensities.}
\label{fig:fig2}
\end{figure}

The simulated probe transmission spectra are given in Fig. \ref{fig:fig1} (b) and (d) for two different pump intensities. There are striking similarities between the experimental and simulated spectra. Firstly, Fig. \ref{fig:fig1} (b) features a long lasting ($\sim \hbar/\Gamma_x$) blue-shift of the lower-polariton in the positive delay, while the blue-shift builds up on a shorter time scale ($\sim 2\hbar/(\gamma_{c}+\gamma_x^*)$) in the negative delay. On the other hand, the energy shift of the upper-polariton is almost zero because of the cancellation of the blue and red-shift contributions induced by $g_0$ and $g_{\rm pae}$ respectively. The high-density simulation (Fig. \ref{fig:fig1} (d)) reproduces both the occurrence of three peaks at negative delays and of a single peak at positive delays. At negative delays, the signal shows a three peak structure, this is because the signal is temporally convoluted due to the finite lifetime of the probe pulse in the sample. Therefore the side peaks arise from the portion of the probe which is transmitted before the arrival of the pump pulse; however, the middle peak builds with the arrival of the pump pulse in the short lifetime of the polarization given by $\sim 2\hbar/(\gamma_c+\gamma_x^*)$. At positive delays, since the probe polarization is always affected by the long lasting pump population, the single middle peak remains for a long time ($\sim \hbar/\Gamma_x$). If the pump intensity is further increased, eventually the term $\gamma_x(N)$ becomes comparable to the effective Rabi coupling $\Omega-2g_{\rm gpae}N$ and the central peak asymptotically reaches the cavity mode $\epsilon_c$, which is the signature of a strong to weak coupling transition \cite{Quochi1998}.

For the purpose of better understanding the effect of the incoherent exciton population and of EID on the two polariton resonances, we present in Fig. \ref{fig:fig2} a simulation without EID or pure dephasing ($g'=\gamma_{x}^{*}=0$ meV). The other parameters are same as for Fig. \ref{fig:fig1}. The probe transmission of Fig. \ref{fig:fig2} is a simulation of the coherent limit, where the polarization decay (dephasing) rate is a half of the population decay rate ($\gamma_x(N)=\Gamma_x/2=0.005$ meV). We find that the polariton branch is broadened towards the high energy side because of ``dynamical energy-shift". Namely, the mean-field energy-shift of polariton temporarily decreases following the decay of the polariton density. The time integration of the temporal decrease of the energy-shift introduces a broadening of the polariton branches. The blue shift of both polariton resonances decay two times faster at positive delays than they emerge at negative ones. In this limit, the dynamics of the exciton population is uniquely determined by the polarization and we can replace the population $N({\bf x},t)$ with the square of the polarization $|P({\bf x},t)|^2$. Here the three sets of equations can be reduced to two equations composed of the exciton polarization and electric field, this is the commonly used exciton-photon GPE \cite{Carusotto2013,comment}. Clearly, the GPE cannot reproduce the dynamics of polaritons in the presence of EID, which is evidenced by the huge differences between Fig. \ref{fig:fig2} and the experiments (Fig. \ref{fig:fig1} (a) and (c)). In particular, the high pump intensity simulation (Fig. \ref{fig:fig2} (b)) reproduces neither the three peak structure nor the disappearance of the quantum beat pattern for the positive delay. This simulation also implies that the bleaching of the upper and lower polariton resonances in the positive delay (Fig. \ref{fig:fig1} (c)) is associated with EID. The transition to the weak coupling regime can also be observed with the GPEs, however the very strong dynamical blue-shift effects completely differ from the observed experimental behaviour.
\begin{figure}
\includegraphics[width=0.48\textwidth]{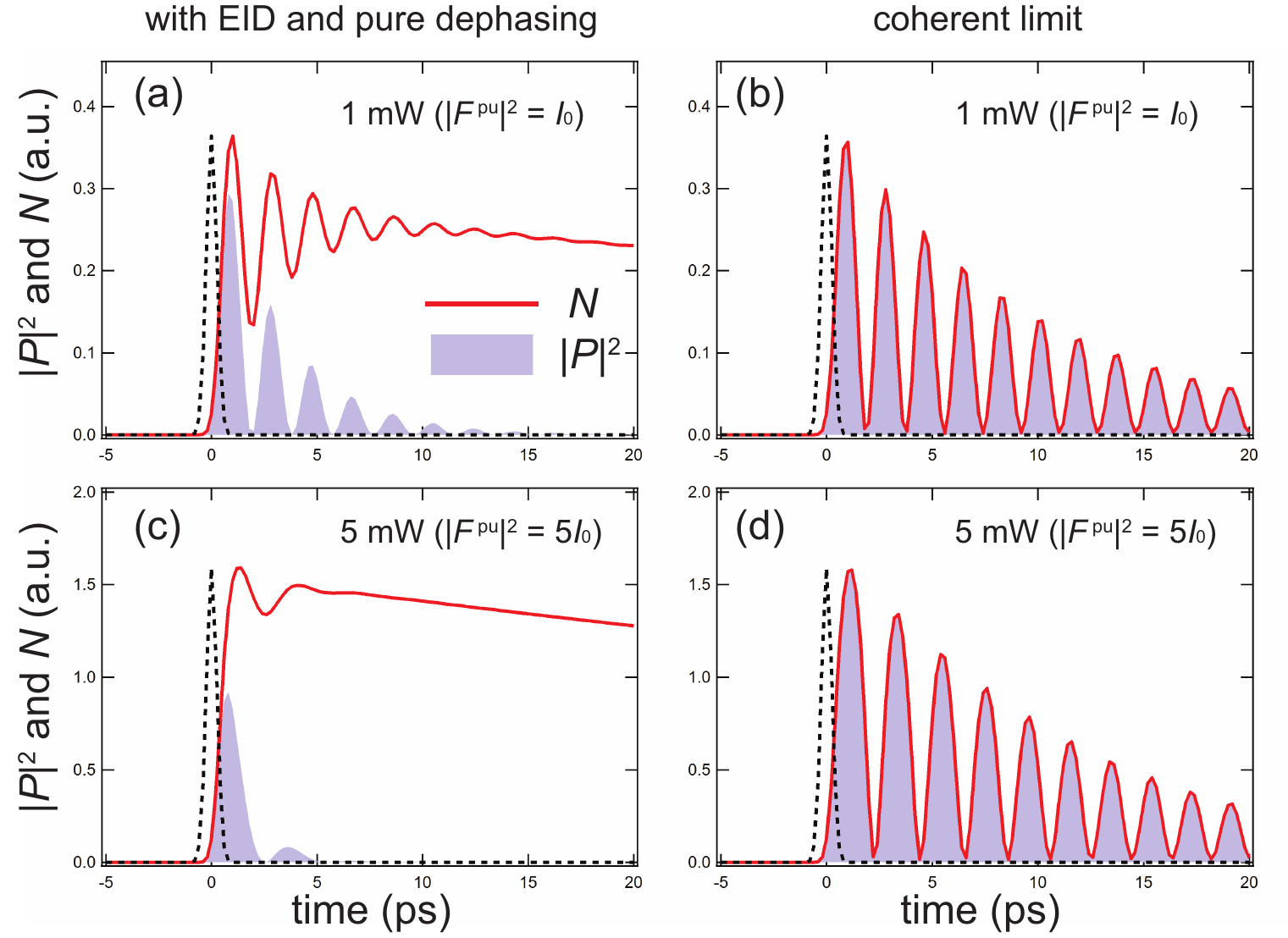}
\caption{(color online) Simulated time evolution of the $\bm k=0$ $\mu$m$^{-1}$ polarization $|P|^2$ and population $N$ as a function of time after an arrival of a single pulse. Simulations with EID and pure dephasing (a,c) and in a coherent limit ($g'=\gamma^*=0$ meV) (b,d) are presented for two different laser pulse intensities 1 mW (a,b) and 5 mW (c,d). The parameters are the same as the simulation of Fig. 1. The dashed lines represent scaled laser pulses.}
\label{fig:fig3}
\end{figure}

In Fig. \ref{fig:fig3}, we present simulated real time evolutions of exciton polarization $P(t)$ and population $N(t)$ at $\bm k=0$ $\mu$m$^{-1}$ after the arrival of a single laser pulse. Without EID (coherent limit), the time evolution of the exciton population $N(t)$ is found to coincide with that of $|P(t)|^2$, which supports the relation $N(t)=|P(t)|^2$ and the factorization $\langle\hat{\bm \psi}_{x}^{\dagger}\hat{\bm \psi}_{x}\rangle=\langle\hat{\bm \psi}_{x}^{\dagger}\rangle\langle\hat{\bm \psi}_{x}\rangle$, which is assumed in the exciton-photon GPE. In this case, the lifetime of the system is mainly determined by the short photon lifetime ($\sim \hbar/\gamma_c$). On the other hand, with EID, we have to distinguish between the dynamics of the polarization and that of the population. While the exciton polarization $P(t)$ is directly coupled to the cavity photon $E(t)$ (not shown), there is no direct coupling between the exciton population $N(t)$ and the photon $E(t)$. Therefore, while the polarization decays with a lifetime of the same order as that of the cavity photon ($\sim \hbar/\gamma_c$), the population decays
independently and stays for a long time ($\sim \hbar/\Gamma_x$), even after the disappearance of the polarization. Intuitively, the EID process can be understood as follows; microscopically, the exciton-exciton collisions introduce an energy fluctuation, which gives an additional random phase to the time evolution of the exciton field operator $\hat{\bm \psi}_{x}({\bf x},0)e^{i(\epsilon_x+\delta\epsilon)t/\hbar}$. Since the phase $\delta\epsilon$ is random, the expectation value $P({\bf x},t)=\langle\hat{\bm \psi}_{x}({\bf x},t)\rangle$ shows a decay \cite{Schmitt-Rink1991}, which is the origin of the imaginary part of the interaction constant $g'$ (EID) \cite{Cohen-Tannoudji1992,Ciuti1998}. Meanwhile, the  the energy fluctuations affect neither the term $\hat{\bm \psi}_{x}^{\dagger}({\bf x},t)\hat{\bm \psi}_{x}({\bf x},t)$ nor its expectation value $N({\bf x},t)=\langle\hat{\bm \psi}_{x}^{\dagger}({\bf x},t)\hat{\bm \psi}_{x}({\bf x},t)\rangle$ due to a phase cancellation \cite{Schmitt-Rink1991}. Finally, we comment that the incoherent exciton population $N$ should be interpreted as the ``inactive" excitonic reservoir already discussed in the context of non-resonantly excited polariton condensates \cite{Lagoudakis2011,DeGiorgi2014}. Actually, both incoherent exciton population featured in Fig. \ref{fig:fig3} and inactive excitonic reservoir have a long lifetime and nevertheless contribute to the energy shift of the polariton resonances \cite{Ferrier2011}.           

In conclusion, we investigated the coherent dynamics of exciton-polaritons by femtosecond pump-probe spectroscopy in a high quality semiconductor microcavity. We demonstrated that excitation-induced dephasing, as a manifestation of exciton-exciton interactions, is a major mechanism for the breaking of the strong coupling regime. We showed that the experimental results could only be simulated with the inclusion of EID in the excitonic Bloch equations. 

The present work is supported by the Swiss National Science Foundation under project N$^{\circ}$135003 and the European Research Council under project Polaritonics contract N$^{\circ}$291120. The polatom network is also acknowledged.

\begin{thebibliography}{35}
\bibitem [{com()}]{comment}%
  \BibitemOpen
  \href@noop {} {\bibinfo  {journal} {Strictly speaking, the coupled-mode
  approximated EBE in coherent limit is slightly different from a coupled-mode
  GPE}\ }\BibitemShut {NoStop}%
\end{thebibliography}

%

\section{Supplemental Material: Coupled mode equations for pump-probe dynamics}
We present explicit forms of the coupled mode equations for simulating pump-probe spectra. We restrict involving modes into three modes: pump, probe and idler. The polarization P, population N and electric field E are respectively written as $N=N^{pu}+N^{pr}e^{ikr}+N^{pr*}e^{-ikr}$, $P=P^{pu}+P^{pr}e^{ikr}+P^{id}e^{-ikr}$, and $E=E^{pu}+E^{pr}e^{ikr}+E^{id}e^{-ikr}$. Substituting these representations into the excitonic bloch equations (EBE) in the manuscript and neglecting components such as $e^{\pm i2k}$ and $e^{\pm i3k}$, we obtain 8 coupled equations of motions. Firstly, the population $N$ read, 
\begin{eqnarray*}
i\hbar \dot{N}^{pu}&=&-i\Gamma_x N^{pu}\\
& &-\Omega a_{pu}+2g_{\rm pae}(a_{pu}N^{pu}+b_{id}N^{pr}+b_{pr}N^{pr*})\nonumber\\
i\hbar \dot{N}^{pr}&=&-i\Gamma_x N^{pr}\\
& &-\Omega b_{pr}+2g_{\rm pae}(b_{pr}N^{pu}+a_{pu}N^{pr}+cN^{pr*})\nonumber\\
\end{eqnarray*}
Now the quantities $a_{pu}$, $b_{pr}$, $b_{id}$ and $c$ are given by 
\begin{eqnarray*}
a_{pu}&=&2i{\rm Im}(P^{pu}E^{pu*}+P^{pr}E^{pr*} + P^{id}E^{id*})\nonumber\\
b_{pr}&=&P^{pu}E^{id*}-P^{id*}E^{pu}+P^{pr}E^{pu*}-P^{pu*}E^{pr}\nonumber\\
b_{id}&=&P^{pu}E^{pr*}-P^{pr*}E^{pu}+P^{id}E^{pu*}-P^{pu*}E^{id}\nonumber\\
c&=&P^{pr}E^{id*}-P^{id*}E^{pr}.\nonumber\\
\end{eqnarray*}
The equations of motion of exciton polarization is written as    
\begin{eqnarray*}
i\hbar \dot{P}^{pu}&=&(\epsilon_x-i\gamma_x)P^{pu}+g(N^{pu}P^{pu}+N^{pr*}P^{pr}+N^{pr}P^{id})\\
& &+\Omega E^{pu}-2g_{\rm pae}(N^{pu}E^{pu}+N^{pr*}E^{pr}+N^{pr}E^{id})\nonumber\\
i\hbar \dot{P}^{pr}&=&(\epsilon_x-i\gamma_x)P^{pr}+g(N^{pu}P^{pr}+N^{pr}P^{pu})\\
& &+\Omega E^{pr}-2g_{\rm pae}(N^{pu}E^{pr}+N^{pr}E^{pu})\nonumber\\
i\hbar \dot{P}^{id}&=&(\epsilon_x-i\gamma_x)P^{id}+g(N^{pu}P^{id}+N^{pr*}P^{pu})\\
& &+\Omega E^{id}-2g_{\rm pae}(N^{pu}E^{id}+N^{pr*}E^{pu}).\\
\end{eqnarray*}
Finally, the electric fields follow the following equations:  
\begin{eqnarray*}
i\hbar \dot{E}^{pu}&=&(\epsilon_c-i\gamma_{c})E^{pu}+\Omega P^{pu}\\
& &-g_{\rm pae}(N^{pu}P^{pu}+N^{pr}P^{id}+N^{pr*}P^{pr}) - f^{pu}_{\rm ext}\nonumber\\
i\hbar \dot{E}^{pr}&=&(\epsilon_c-i\gamma_{c})E^{pr}+\Omega P^{pr}\\
& &-g_{\rm pae}(N^{pu}P^{pr}+N^{pr}P^{pu}) - f^{pr}_{\rm ext}\nonumber\\
i\hbar \dot{E}^{id}&=&(\epsilon_c-i\gamma_{c})E^{id}+\Omega P^{id}-g_{\rm pae}(N^{pu}P^{id}+N^{pr*}P^{pu}).\nonumber\\
\end{eqnarray*}

\end{document}